\documentclass[twocolumn,showpacs,preprintnumbers,amsmath,amssymb,prb,floatfix]{revtex4}  

\usepackage{graphicx}% Include figure files   
\usepackage{dcolumn}% Align table columns on decimal point    
\usepackage{bm}% bold math
\usepackage{mhchem}
\usepackage{epstopdf}
\usepackage{color}
\usepackage{soul} % use this (many fancier options)

\begin{document}

%\title{Pressure Induced Percolative   metal-insulator transition    in LaMnO$_3$  }     
\title{Percolative Metal-Insulator Transition in LaMnO$_3$  }   
\author{M. Sherafati$^{1\dagger}$}
\author{M. Baldini$^{2,3}$}
\author{L. Malavasi$^4$}
\author{S. Satpathy$^1$}
\affiliation{$^1$Department of Physics $\&$ Astronomy,
University of Missouri,
Columbia, MO 65211, USA}
\affiliation{$^2$Geophysical Laboratory, Carnegie Institution of Washington, 5251 Broad
Branch Rd, NW Washington, DC 20015, USA}
\affiliation{$^3$HPSynC, Geophysical Laboratory, Carnegie Institution
of Washington, Advanced Photon Source, Argonne National Laboratory, 9700
South Cass Avenue Argonne,
IL 60439, USA}

%\affiliation{$^3$Geophysical Laboratory, Carnegie Institution of Washington, Washington, D.C. 20015, USA}

\affiliation{$^4$Department of Chemistry and INSTM, University of Pavia, Viale Taramelli 10-16, Pavia, Italy}

\date{\today}

\begin{abstract}
We show 
that the pressure-induced metal-insulator transition  (MIT) in LaMnO$_3$ is fundamentally different from the Mott-Hubbard transition and is percolative in nature, with the measured resistivity obeying the percolation scaling laws. 
Using the Gutzwiller method to treat correlation effects in a model Hamiltonian that includes both Coulomb and Jahn-Teller interactions,
we show, One, that the MIT is driven by a competition between electronic correlation and the electron-lattice interaction, an issue that has been long debated, and Two, that with compressed volume,
the system has a tendency towards phase separation into insulating and metallic regions, consisting, respectively,  of Jahn-Teller distorted and undistorted octahedra. This tendency manifests itself in a mixed phase of intermixed insulating and metallic regions in the experiment. Conduction in the mixed phase occurs by percolation 
 and the MIT occurs when the metallic volume fraction, steadily increasing with pressure, exceeds the percolation threshold $v_c \approx 0.29$. Measured high-pressure resistivity follows the  percolation scaling laws quite well, 
 % establishing the percolative nature of conduction,
and the temperature dependence  follows the Efros-Shklovskii variable-range hopping behavior  for granular materials. 

% OLD-- We show  that the pressure-induced metal-insulator transition  (MIT) in LaMnO$_3$ is fundamentally different from the Mott-Hubbard transition and is percolative in nature, with the resistivity obeying the percolation scaling laws.    Using the Gutzwiller method to treat correlation, we find that the MIT is driven by a competition between electronic correlation and the electron-lattice interaction.    With applied pressure, an inhomogeneous phase of intermixed insulating and metallic regions, consisting, respectively,  of Jahn-Teller distorted and undistorted octahedra,  develops and the MIT occurs when the metallic volume fraction exceeds the percolation threshold $v_c \approx 0.29$. High-pressure measurements establish the percolative power-law scaling for the resistivity   and the temperature dependence follows the Efros-Shklovskii variable-range hopping behavior  for granular materials. 

\end{abstract}
\maketitle

\section {Introduction} 
The doped manganites such as La$_{1-x}$Ca$_x$MnO$_3$   
%showing the much studied  colossal magetoresistance effects 
are unique systems for studying competing interactions between spin, electronic, orbital, and lattice degrees of freedom.\cite{Dagotto, Tokura, Satpathy2}  
The end member LaMnO$_3$ (LMO) is of special interest, since, while being governed by  the same  interactions, it is at the same time free from clutter due to the Ca dopants.  
The  pressure induced metal-insulator transition  (MIT) in LMO  has been long debated.  
There are two issues.
First, while resistance measurements indicate a sharp transition to the metallic state at the critical pressure $P_c \approx 32$ GPa \cite{Loa}, Raman measurements, on the other hand, show a gradual change with both Jahn-Teller (JT) distorted and undistorted regions persisting over a wide range of pressure \cite{Loa, Baldini, Ramos}.  An understanding of the MIT must explain this dual behavior, which we explain below in terms of percolation.

The second issue is the relative role of the competing interactions
%, in particular the electron-electron  vs. the electron-lattice interaction, 
in mediating the MIT.  
%Regarding the driving mechanism of the MIT, there is no general consensus.   
Loa \textsl{et al.} \cite{Loa} first suggested that the MIT is driven by band-width ($W$) enhancement with pressure, based on the fact that the JT distortion disappears  
%at 18 GPA,  which is much below $P_c$. 
much below $P_c$ and therefore has no role to play, so that the change in $U/W$ results in an MIT of the standard Mott-Hubbard type\cite{Mott}. 
This conclusion was refuted by Baldini and Ramos and coworkers\cite{Baldini, Ramos}, who observed, to the contrary, that the   distortions in fact persist beyond the MIT and remain relatively unchanged across the transition. 
%In addition, the observation of a mixed phase of distorted and undistorted octahedra on either side of the MIT led to the conclusion that it is not a standard Mott-Hubbard insulator.  
%
%
Several  theoretical studies \cite{Trimarchi,Yamasaki,Yin, Fuhr, Koch} also suggested that both the Coulomb as well as the JT  interaction are important questioning the pure Mottness of the observed MIT. 
Trimarchi and Binggeli\cite{Trimarchi} studied the Mn-O distances under pressure with the Coulomb-corrected LDA+U density-functional method and found the Coulomb interaction to be essential in establishing the insulating ground state.
% of the LMO \textsl{Pnma} structure.
Based on the dynamical-mean-field results (LDA+DMFT), Yamasaki \textsl{et al.} \cite{Yamasaki} argued that both the JT and the Coulomb interactions are important for the MIT. 
Similar conclusion was found from the slave-boson solution of a model Hamiltonian\cite{Fuhr}.
Yin \textsl{et al.} \cite{Yin} suggested that the JT distortion is facilitated by the Coulomb U term via enhanced localization. 
%Using the slave-boson approach for a model Hamiltonian, Fuhr \textsl{et al.} \cite{Fuhr} found that the JT distortion vanishes only after the metallic phase is entered and both Coulomb and JT interactions are necessary for the MIT. 
Considering another aspect of the problem,
Koch \textsl{et al.} \cite{Koch} showed that in order to describe the orbital ordering  seen in neutron scattering, the JT interactions are important, and the Kugel-Khomskii superexchange derived from the Coulomb U term is not sufficient for it. 
Much of this theory work was aimed at the understanding of the role of the competing interactions, rather than the phase coexistence across the MIT, although a recent hybrid-functional calculation\cite{He} found different magnetic phases to be close in energy at T = 0,
suggesting the propensity towards phase coexistence.
%which however does not necessarily imply phase coexistence. 

%Mohammad: Nonetheless, there is no theoretical explanation for the phase coexistence also observed at room temperature. Moreover, to this date, not only no theoretical interpretation has ever been made for the pressure and temperature dependence of the LMO resistance in the insulating regime reported originally in Ref. [\onlinecite{Loa}] but also there is no such data for the metallic regime beyond the MIT pressure. The overarching goal of our work is to demonstrate how percolation theory forms the bedrock of the underlying physics of the MIT in  LMO.

In this paper, from a Gutzwiller solution of a model Hamiltonian and high-pressure transport measurements, we show that the pressure-induced MIT in undoped LMO is percolative in nature. 
In other words, conducting transport does not occur as a result of the formation of a homogeneous metallic phase, as happens in the Mott-Hubbard MIT, but rather, it occurs when the volume fraction of the metallic region, gradually increasing with pressure, exceeds the percolation threshold. 
 The overarching goal of our work is to demonstrate how percolation theory forms the foundation of the underlying physics of the MIT in  LMO.
 We focus on the high-temperature paramagnetic phase, so that the transport is uncluttered by the magnetic transitions that exist at low temperatures.

\section{Model Hamiltonian and Gutzwiller Solution}  

We consider a two-band, spinless model Hamiltonian, containing the key Coulomb and  JT interactions:
\begin{align}
{\cal H}=&\sum_{\langle ij \rangle, \alpha \beta}  t^{\alpha \beta}_{ij}   (\hat c^\dag_{i\alpha}\hat c_{j\beta}+\text{H.c.})
- g  \sum_i (Q_{i3}\hat{\sigma}_z+Q_{i2}\hat{\sigma}_x)  \nonumber \\
+&\frac{1}{2}K\sum_i(Q_{i3}^2+Q_{i2}^2)+U\sum_i \hat n_{i1}\hat n_{i2},
 \label{eq:Hamil}
\end{align}
where $\hat c_{i\alpha}^\dagger$ creates an $e_g$ electron in orbital $\alpha$ ($=1,2$) at site $i$ on the simple cubic Mn lattice, 
%\clr{with lattice parameter $a$}, 
$\vec \sigma$ is the pseudospin describing the two $e_g$ orbitals, $|\uparrow\rangle =| x^2-y^2\rangle$ and $|\downarrow\rangle =| z^2\rangle$, $Q_2$ and $Q_3$ are the two
JT distortion modes of the MnO$_6$ octahedron, and  
$K$ and $U$ are the elastic constant for the JT modes and the intra-orbital on-site Hubbard U Coulomb interaction, respectively.
Only one spin is included in the Hamiltonian due to the following reason.
Because of the large Hund's coupling $J_H \rightarrow \infty$, the $e_g$ spins are always parallel to the core $t_{2g}$ spins, with the result that the antiparallel spin states are altogether omitted
due to their high energy. The two $e_g$ states 
in the Hamiltonian, Eq. \ref{eq:Hamil}, therefore have their spins aligned with the local core spin, which can however vary from site to site.

Although the $t_{2g}$  core spins are not explicitly included in the Hamiltonian, their effect on the hopping of the $e_g$ electrons is a crucial part of the physics of the manganites and must be taken into account. 
%Because of the large Hund's coupling $J_H \rightarrow \infty$, the $e_g$ spins are always parallel to the core $t_{2g}$ spins, with the result that the antiparallel spin states are altogether omitted due to their high energy. The two $e_g$ states  in the Hamiltonian, Eq. \ref{eq:Hamil}, therefore have their spins aligned with the local core spin, which can vary from site to site. 
%
The core spins modify the hopping integrals between the $e_g$ electrons, since they are always aligned parallel with the core spins on each lattice site,   via the Anderson-Hasegawa double exchange by the factor $\cos(\theta /2)$, where $\theta$ is the angle between  two neighboring core spins, treated as classical\cite{Anderson-Hasegawa}. 
As we are interested in the paramagnetic phase at room temperature, 
the random thermal fluctuations lead to the random fluctuations of the orientation of the core spins at each site, so that the thermal   average of the Anderson-Hasegawa factor
yields the result, $\langle \cos(\theta /2) \rangle = 2/3$, which modifies the hopping integral between the $e_g$ electrons.

%Since the $t_{2g}$  core spins are not explicitly included in the Hamiltonian, they modify the hopping integrals  by the well known Anderson-Hasegawa double exchange factor $\cos(\theta /2)$, where $\theta$ is the angle between two neighboring core spins\cite{Anderson-Hasegawa}. 

%

%
To describe the effect of pressure, we take the hopping integral to be volume dependent with  $t^{\alpha \beta} (r) \propto r^{-7}$ following  Harrison scaling\cite{Harrison}, add a Madelung term $E_M$
and a repulsive interaction term $E_R$ between the ions to keep the crystal from collapsing. 
The total energy then becomes   
$
E =  E_{el}  +E_M + E_R,
$
and we have  used the simplified forms $E_M = -A/r$, $E_R = B/r^{12}$, and as usual, $t^{\alpha \beta}$ may be expressed in terms of the $dd\sigma$ hopping integral, denoted here by  $-t$. Guided by the literature\cite{Satpathy1, Satpathy2, Millis, Tang}, we set the  parameters $A = 6$ eV, $B = 0.5$ eV, $g = 2.5$ eV/ \AA, $K = 10$ eV/\AA$^2$,  $U = 3$ eV, and $t = 0.6$ eV.

% which describe the structural and electronic properties of LMO reasonably well \cite{Satpathy1, Satpathy2, Millis}. \clr{The Madelung and repulsive energy parameters in our model serve the purpose of pronounced demonstration of the phase separation in the total energy. We have checked that using realistic parameters \cite{Tang} still preserves this feature.}  
%Inclusion of these terms allows us to model the total energy of the distorted and undistorted JT phases as a function of pressure.

%: Gutzwiller
We have solved the model using the Gutzwiller approximation for the Coulomb interaction term
in Eq. \ref{eq:Hamil}, treating the two $e_g$ orbitals as pseudo-spins. 
The Gutzwiller wave function is given by
\begin{align}
\vert\Psi_G\rangle=\eta^{\hat D}\vert\Psi_0\rangle,
% \label{eq:GWF}
\end{align}
where $\vert\Psi_0\rangle$ is the uncorrelated many-body wave function, $\hat D$ counts the site double occupancy,  and the Gutzwiller variational parameter $\eta$ is obtained by minimizing the expectation value of energy $\langle\Psi_G\vert{\cal H}\vert\Psi_G\rangle$. 
In the thermodynamic limit,  the average double occupancy $ d \equiv \langle \hat D \rangle$ is related to $\eta$ by the expression $\eta^2 = 4 d^2 [(1-2d)^2 - m^2]^{-1}$. The electrons hop in a correlated manner, leading to a reduced kinetic energy, described by the Gutzwiller reduction factor 
\cite{Gutzwiller, MSashiGutzwiller}
\begin{align}
\gamma(m,d)=\frac{2d\left(\sqrt{1-m-2d}+\sqrt{1+m-2d}\right)^2}{1-m^2},    
\label{eq:Gamma}
\end{align}
valid for the half-filled case, viz.,  $n_1 + n_2 = 1$ (one $e_g$ electron per site), with $m=\langle \hat n_{2}-\hat n_{1}\rangle$ being the orbital polarization. A small $d$ as compared to the uncorrelated value $d_{\text{uncorr.}} = n_1 n_2$ indicates a strongly correlated state
and according to the Brinkman-Rice criterion\cite{Brinkman}, a  Mott-Hubbard insulating ground state is indicated if $d \rightarrow 0$.

The band structure energy is computed by taking into account this reduction factor and diagonalizing the $2 \times 2$ Bloch Hamiltonian in the orbital space
\begin{align}
H_k= 
\left( \begin{array}{cc}
  \varepsilon_{11} (k) - gQ_3&        \varepsilon_{12} (k)  - g Q_2\\ 
\varepsilon_{12} (k)  - g Q_2  & \varepsilon_{22} (k)  + g Q_3
\end{array}\right),
 \label{eq:Hk}
\end{align}
where $\varepsilon_{11}( k)= \bar V  (\cos k_xa+\cos k_ya+4\cos k_za)/2 $, 
$\varepsilon_{12}( k)= -\sqrt {3} \bar V (\cos k_x a-\cos k_y a) / 2 $, 
$\varepsilon_{22} ( k)=   3 \bar V (\cos k_xa+\cos k_ya)/2 $, 
and $\bar{V}   =  - (2/3)   \gamma(m,d)  t (r)$, 
with $t (r) \propto r^{-7}$ and the factor 2/3 coming from the Anderson-Hasegawa
renormalization as already discussed.
We minimized the total energy per lattice site
\begin{equation}
E = \sum^{\text{occ}}_{\bm k\nu}\varepsilon_{\bm k \nu }(d,Q_2, Q_3)+\frac{1}{2}KQ^2+Ud    
+E_M+E_R,
\label{eq:Etot}
\end{equation}
as a function of $d$ and $Q_i$ for each volume, which yields the ground-state solution. Here $Q \equiv (Q_2^2 + Q_3^2)^{1/2}$ and $\varepsilon_{\bm k \nu }$ are the band structure energies obtained by diagonalizing the Hamiltonian $H_k$, Eq. \eqref{eq:Hk}.

%:Fig. 1
\begin{figure}[t]
%\begin{figure}[!htb]
%\includegraphics[angle=0,width=0.75 \linewidth]{Maxwell.pdf}   
\includegraphics[angle=0,width=0.8 \linewidth]{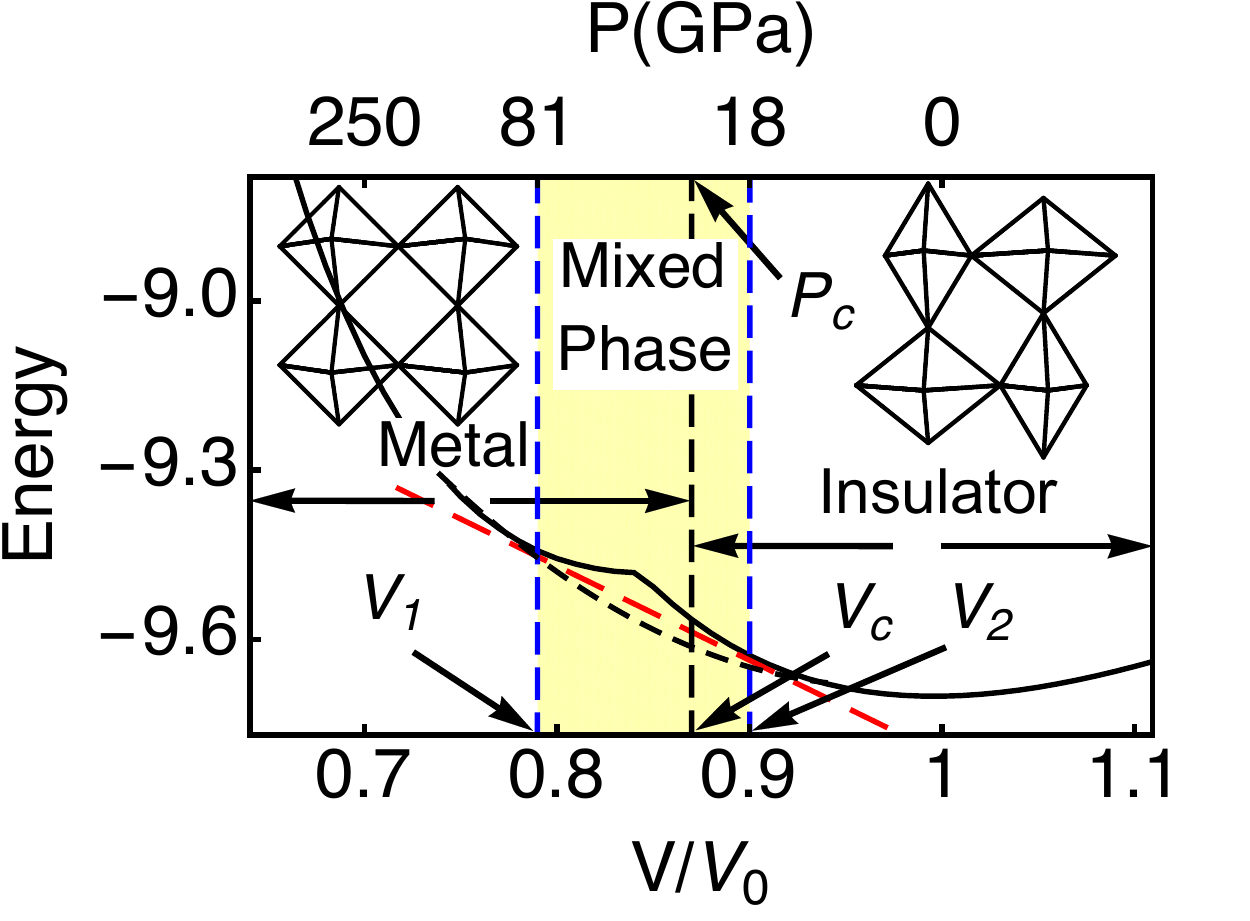}
\caption{ (Color online) Total energy as a function of volume obtained from Eq. \eqref{eq:Etot} for parameters corresponding to LMO, indicating  regions of JT distorted and undistorted phases. 
%and a propensity towards the formation of a mixed phase. 
As volume is compressed below $V_2$, a metallic component begins to form, and the system conducts below $V_c$ (black dashed line),
when the metallic volume fraction $v$, calculated from the  Maxwell-construction result, Eq. (\ref{vm}), 
%using the computed threshold volumes $V_1$ and $V_2$,
exceeds the percolation threshold $v_c \approx 0.29$. The  corresponding threshold pressure for MIT  is $P_c \approx 31$  GPa as computed from the 
measured equation of state\cite{Loa}. 
%The volume fraction $v$, which varies with the total volume $V$, is obtained from the Maxwell-construction result, Eq. (\ref{vm}), using the computed threshold volumes $V_1$ and $V_2$.
Energy is in units of $t$ and volume is in units of $V_0$, the zero-pressure volume. 
The black dashed line (schematic) indicates the mixed phase region, if the phase separation   is suppressed either due to interaction between the phases or for kinetic reasons (see text). 
} 
\label{fig:EvsV} 
\end{figure}

%:Fig. 2
\begin{figure}[b]
\includegraphics[angle=0,width=0.85 \linewidth]{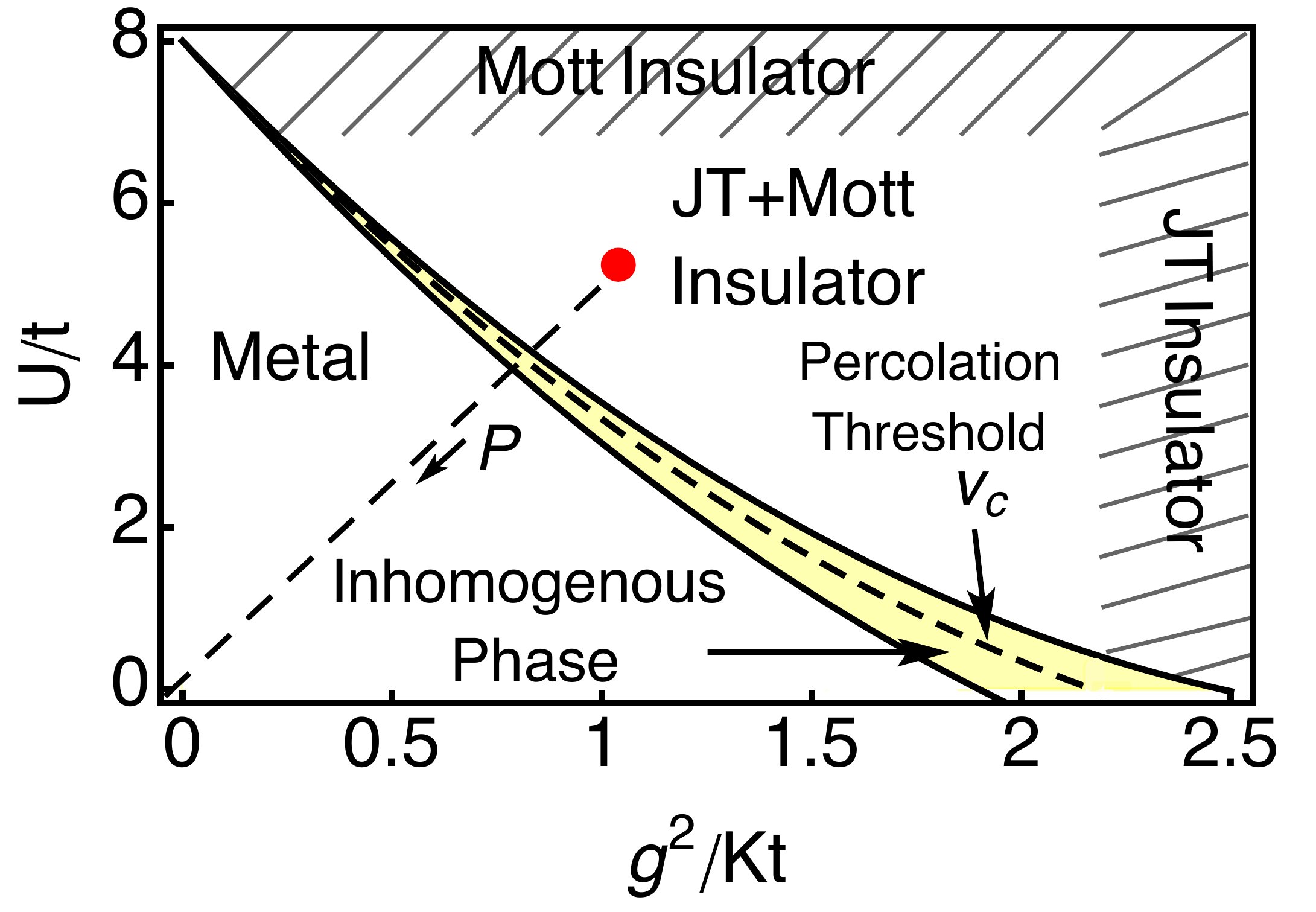}   
\caption{ (Color online) Phase diagram showing the 
% competition between Coulomb and JT interactions. 
metallic and insulating regions, bridged by the inhomogeneous phase (shown in yellow).  The system LMO, starting with the
red dot at ambient pressure, moves along the dashed line as pressure is applied, first entering the inhomogeneous phase while still maintaining its insulating character, until the metallic  fraction exceeds the percolation threshold $v_c$ (curved dashed line).
Finally,  it crosses over to the fully metallic phase, where the metallic domains fill the entire volume. 
%The mixed phase region was constructed by taking the locus of the parameters corresponding to volumes $V_1$, $V_2$, and $V_c$ in Fig. \ref{fig:EvsV}.
} 
\label{PhaseDiagram} 
\end{figure}

\section{Theory Results} 

The total energy, calculated from Eq. (\ref{eq:Etot}), is plotted in Fig. \ref{fig:EvsV} for parameters corresponding to LMO as discussed earlier.
 It shows a double minimum  as a function of volume corresponding to a JT distorted and an undistorted phase, indicating a phase separation in a range of volume, shaded yellow in the figure. For volume constrained in the shaded region, the double minimum would imply the coexistence of two different phases, a high-volume insulating phase with volume $V_2$, and a low-volume metallic phase with volume $V_1$, with a sharp boundary between them. 
 If pressure is fixed, then a first-order transition from the insulating to a metallic phase at a pressure corresponding to the common tangent would be implied. In the experiments, such a sharp transition is, however, not observed. For example, 
 the equation of state shows a continuous change of volume with pressure.\cite{Loa}
 
 The reason for the mixed phase, ubiquitous in the manganites, rather than a phase separation is a topic of considerable interest.
 A phase separated system could be energetically unfavorable due to multiple reasons, not included in our model. For example, presence of a small amount of charged impurities because of unintentional doping could cause a 
 deviation from charge neutrality of the two components and would impede the formation of the phase separation due to the large cost in Coulomb energy. 
 It would instead lead to
a nanoscale inhomogeneous phase (or mixed phase) with intermixed metallic and insulating components   (Coulomb frustrated phase separation)\cite{Bangalore}. It has also been suggested that the mixed phase could even originate due to kinetic reasons, i.e., self-organized  inhomogeneities resulting from a strong coupling between electronic and elastic degrees of freedom\cite{Ahn}. 
 %
%However, in practice, the system does not show a clean phase separation with a sharp boundary between the two phases, but rather a nanoscale inhomogeneous phase (or mixed phase) with intermixed metallic and insulating components.
%This could originate from the Coulomb interaction between the two components\cite{Bangalore} due to deviation from charge neutrality because of unintentional doping (Coulomb frustrated phase separation) or it could originate due to kinetic reasons, i.e., self-organized  inhomogeneities resulting from a strong coupling between electronic and elastic degrees of freedom\cite{Ahn}.

In fact, a number of experiments point to the existence of the mixed phase in LMO under pressure. These experiments include
the   Raman measurements\cite{Ramos, Baldini}, the continuous equation of state \cite{Loa}, as well as the present transport measurements. Of these, the Raman and the high-pressure resistivity measurements show that
the metallic component slowly grows with pressure, while the equation of state indicates that no abrupt volume change occurs with pressure, which is consistent with the existence of the mixed phase.  
Even though the metallic fraction slowly grows with pressure, the transition to metallic conduction is, nevertheless, still sharp and occurs when the metallic fraction exceeds the percolation threshold.  

The metallic fraction may be obtained from the Maxwell construction (red dashed line in Fig.  \ref{fig:EvsV}). 
If $f_1 (f_2)$ is the fraction of the substance in metallic (insulating) phase  in the mixed phase region ($V_1 < V < V_2$), $V$ being the total volume, 
then we have the two equations: $ f_1 + f_2 = 1$ and $f_1 V_1 + f_2 V_2 = V$, solving which we find the volume fraction of the metallic phase
\begin{equation}
v \equiv \frac{f_1 V_1} {V} = \frac{V_2/V-1}   {V_2 / V_1 -1 }.
\label{vm}
\end{equation}
The MIT occurs, when $v > v_c \approx 0.29$, the percolation threshold, 
when the metallic regions begin to touch and percolative conduction begins.
We readily find from Eq. (\ref{vm}), the threshold volume for metallic conduction  $V_c = (v_c / V_1 + (1-v_c) / V_2  )^{-1}$
and the corresponding  $P_c$ is found from the measured equation of state\cite{Loa}, and both are shown in Fig. \ref{fig:EvsV}.

%:Fig. 3
\begin{figure}[t]
%\begin{figure}[!htb]
\hskip - 5 mm
\includegraphics[angle=0,width=0.70 \linewidth]{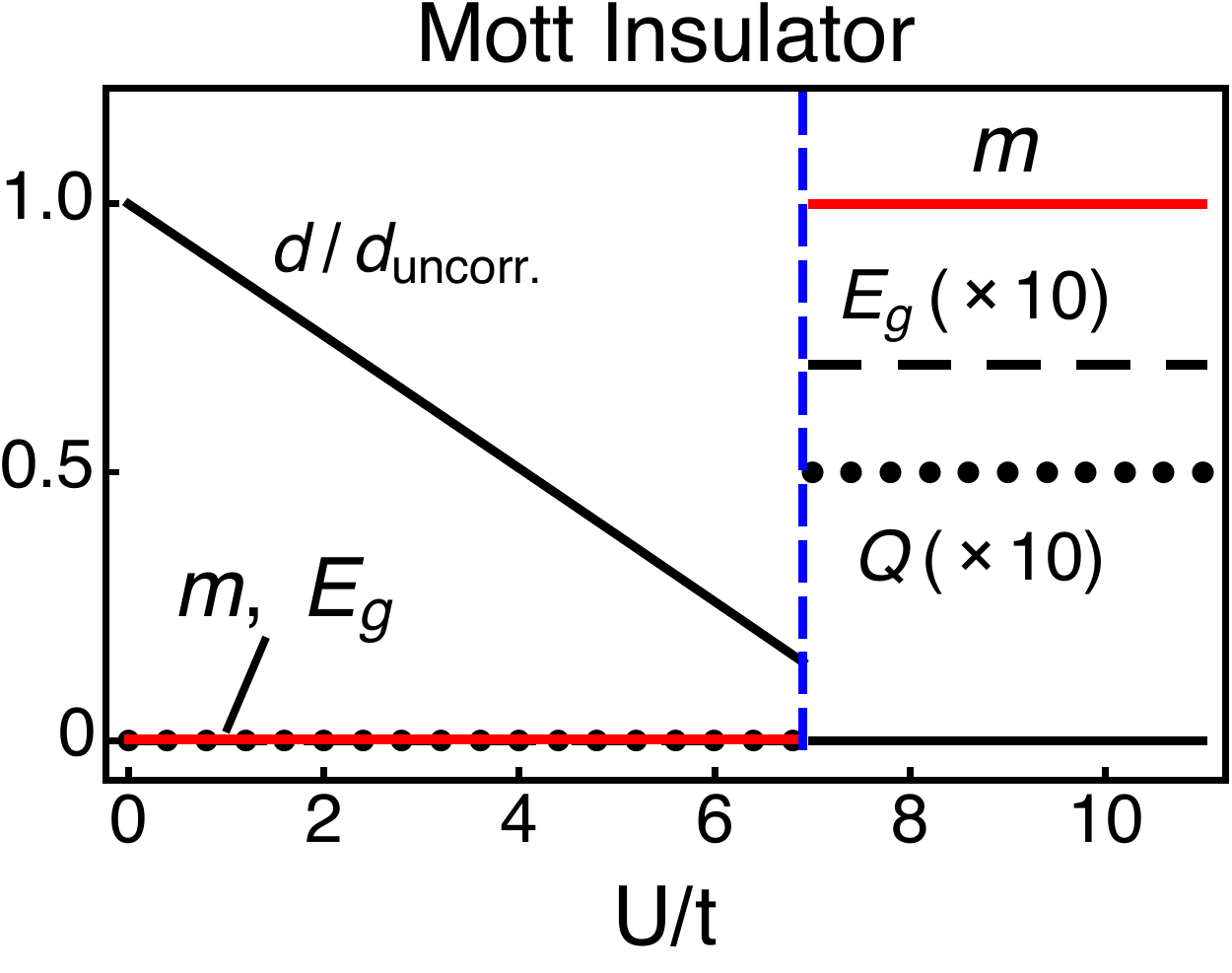}
\vskip5mm
\includegraphics[angle=0,width=0.70 \linewidth]{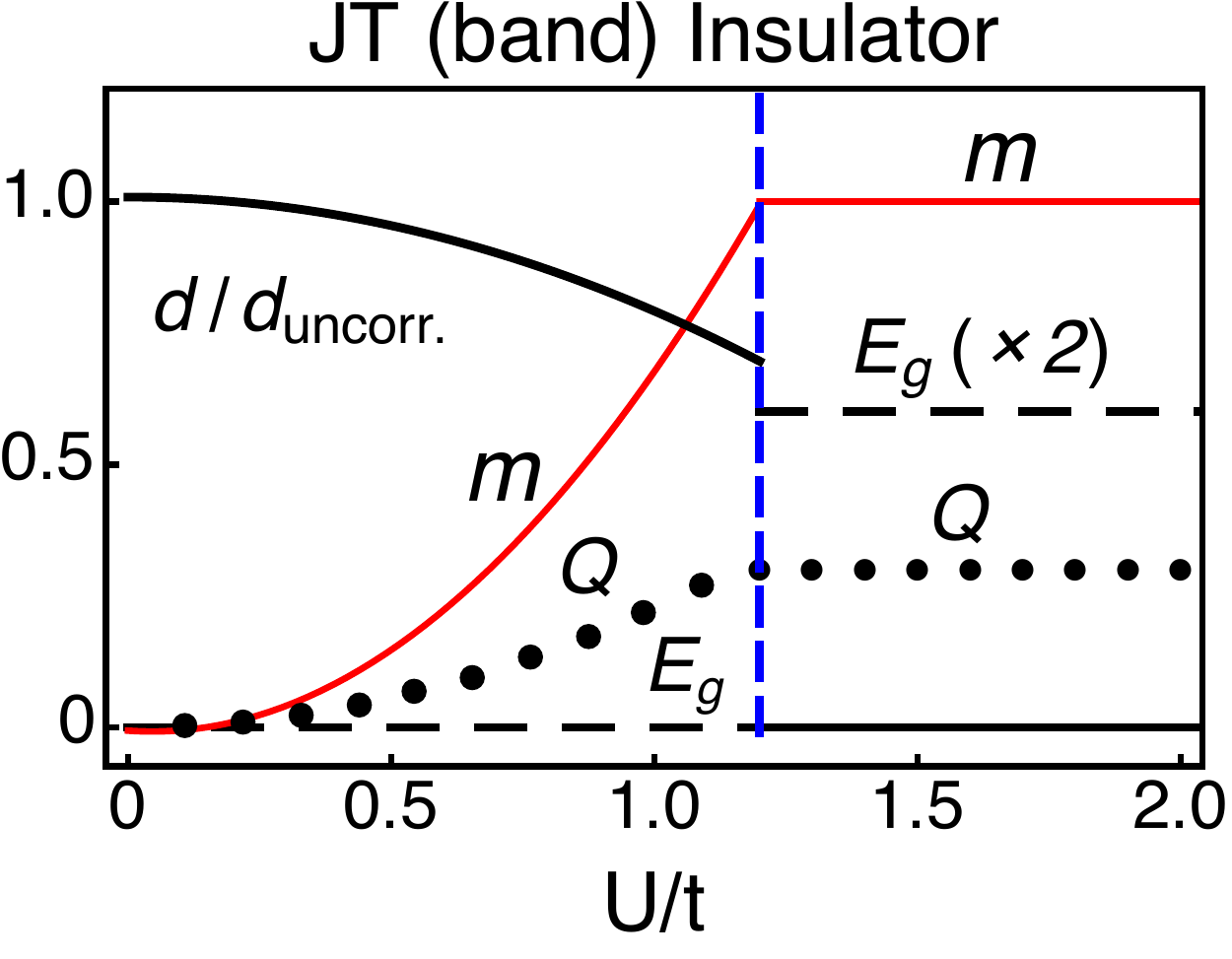}
\caption{(Color online) Contrasting Mott insulator vs. JT band insulator. For weak JT coupling, $g^2/Kt = 0.04$ ({\it top}), the MIT is correlation-driven with the Gutzwiller double occupancy taking the Brinkman-Rice value $d \approx 0$ at the transition point, while  in the opposite, strong-coupling limit, $g^2/Kt = 1.5$ ({\it bottom}), the MIT is driven by a large JT distortion $Q$, with $ d$ hardly changed from its uncorrelated value.  
In this figure, the system is assumed to be always in the homogeneous phase, so that the MIT corresponds to the kink in total energy like in Fig. \ref{fig:EvsV} and not to the percolative MIT.  Here, distortion $Q$ is in \AA, gap $E_g$ is in units of $t$, $m$ is the orbital polarization, and  left of the blue line is a metal, while the right of it is an  insulator.
} 
\label{fig:MottvsJT}
\end{figure}
%

%We have studied the nature of the MIT over a wide range of parameters and the results are  summarized in the phase diagram Fig. \ref{PhaseDiagram}. 
Fig. \ref{PhaseDiagram} summarizes the phase diagram, illustrating the competition between the Coulomb and the JT interactions. The phase diagram was calculated by starting with a fixed parameter set $U$, $g$, and $t$, e.g., the red dot in Fig. \ref{PhaseDiagram} corresponds to LMO at ambient pressure, and then by changing volume which scales these parameters.
With decreasing volume (increasing pressure), the hopping integral $t$ increases much more rapidly as compared to the other parameters
 (taken to be volume independent in our model), so that the system moves along the dashed line towards the origin as shown in the figure (if $t$ doubles, then both $U/t$ and
 $g^2/(Kt)$ are halved). As the system traverses along the line, the volume changes and with it, the total energy, as shown in 
 Fig. \ref{fig:EvsV}, from which the boundary of the inhomogeneous phase and the percolation threshold are determined.
Fig. \ref{PhaseDiagram} was obtained by studying the system traversing along a series of such lines in the parameter space.

The phase diagram, Fig. \ref{PhaseDiagram}, shows distinct behaviors in different regions of the parameter space, viz., metallic behavior, insulating behavior driven by either correlation or Jahn-Teller interaction, or a mixed phase in the crossover region between the metal and insulator. For large Coulomb interaction, one gets a Mott-Hubbard insulator, while for a large JT coupling, one obtains a JT band  insulator  as a large gap opens up between the two orbitals due to a strong JT splitting.

The contrast between the Mott and the JT band insulator is illustrated in Fig. \ref{fig:MottvsJT}, where we have shown the change of the various quantities as the transition point is crossed.
When $g$ is zero or close to zero, we get the standard Mott-Hubbard MIT, in the sense that there is an abrupt change from the metallic state to the insulating state as $U/t$ is increased beyond a critical value, and the system always remains in a single phase, either metallic or insulating. The Gutzwiller double occupancy $d$ is zero at the MIT point, following the Brinkman-Rice criterion\cite{Brinkman}. 
On the other hand, if $g$ is strong as compared to $U$, then correlation effects become negligible, 
and the MIT occurs because $Q$ becomes large and the gap opens up because the energy separation between the two $e_g$ orbitals,
 $2 g Q$, becomes larger as compared to the band width, 
 leading to a JT band insulator. In this case,  $d$ does not change very much from its uncorrelated value as the MIT point is approached. 
 At ambient pressure, LMO is in an intermediate regime, where the insulating state is formed by a combined effect of both Coulomb as well as JT interactions, as indicated by the red dot in Fig. \ref{PhaseDiagram}.
\begin{figure}[b]
\includegraphics[angle=0,width=0.8 \linewidth]{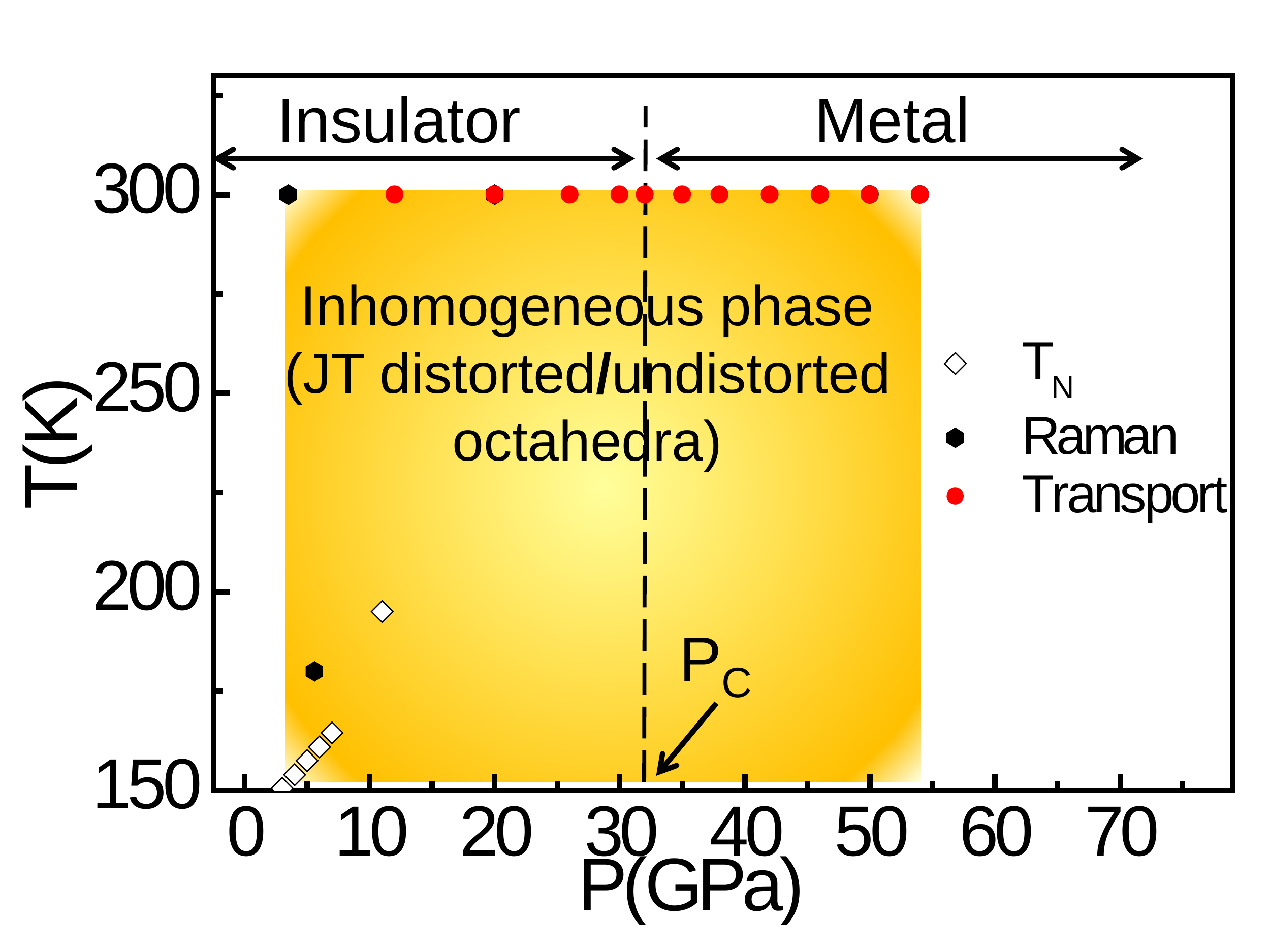}
\caption{(Color online) Summary of the experimental high-temperature phase diagram. The observed inhomogeneous phase region is shaded yellow. The measured resistance corresponding to the red dots are shown in Fig. \ref{fig:scaling} .
} 
\label{Phase} 
\end{figure}

\section{Transport Measurements and Percolation Laws}

We have studied the mixed phase region experimentally from high-pressure transport measurements, which clearly shows the transport behavior characteristic of an
inhomogeneous (or mixed) phase with intermixed metallic and insulating regions. 
We measured the electrical resistance across the metal-insulator transition  region as a function of temperature and pressure up to 54 GPa. 

In our experiments, samples of LMO were synthesized by solid-state reaction starting from 99.999 \% pure La$_2$O$_3$  and Mn$_2$O$_3$ and the oxygen stoichiometry was confirmed by thermo-gravimetric analysis. For the transport experiments, a miniature non-magnetic diamond anvil cell  was employed together with a Re gasket, previously insulated.  The LMO powder was loaded in a 70 micron hole and four platinum leads (2 micron thick) were placed in electric contact with the sample to measure resistance in quasi-four probe configuration using PPMS. At each pressure, 
 resistance data were collected over cooling and warming temperature cycles (10-300 K) \cite{Expt-paper}.
Pressure was measured using the ruby fluorescence technique. 
The resistance changed by   five orders of magnitude as the pressure was varied across the MIT transition  occurring at $P_c \approx 32 - 35$  GPa.

Fig. \ref{Phase} summarizes the high-temperature phase diagram, focusing on the paramagnetic region, which we have studied in the present work. The figure was constructed on the basis of the current experiment and earlier Raman\cite{Baldini}  and N\'eel-temperature measurements.\cite{Zhou}

\begin{figure}[t]
\includegraphics[angle=0,width=0.8 \linewidth]{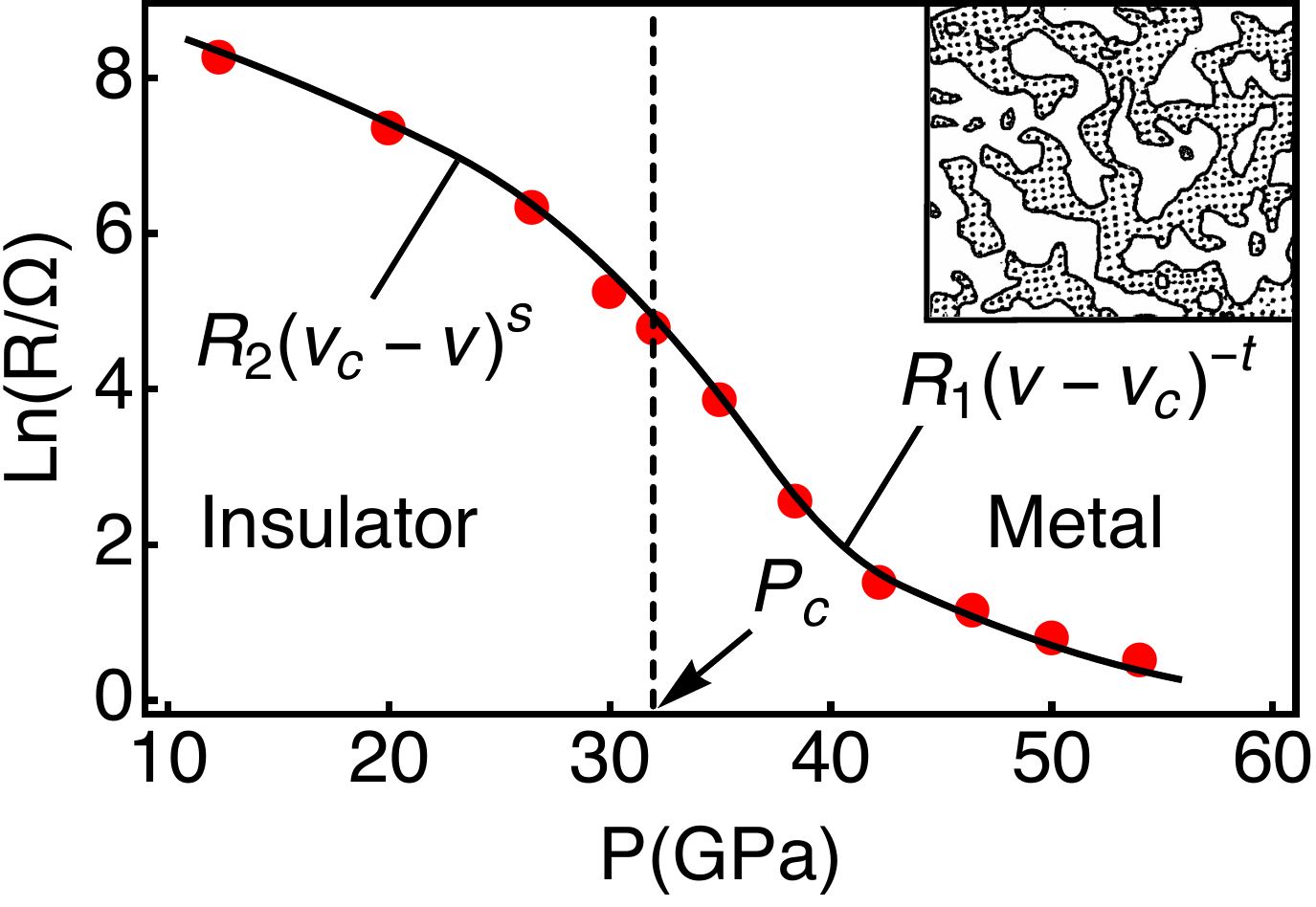}
\caption{(Color online) Measured resistance in the paramagnetic phase ($T=300 K$)  as a function of pressure showing percolative conduction in the  mixed phase region. 
Close to the MIT, the resistance  follows the percolation scaling laws, Eq. \ref{critical},  with the critical exponents $t = 2.1 \pm 0.2 $ and $s =0.9  \pm 0.2$ and the  fitted  resistance constants $R_1 = 0.19\  \Omega$ and $R_2 = 5840 \ \Omega$ (solid curve is a guide to the eye).  
The {\it inset} is a schematic of the inhomogeneous phase near the percolation threshold.  
The metallic volume fraction $v$ was calculated 
by first computing the volume $V$ for a given pressure from the experimental equation of state\cite{Loa} and then
using the  
Maxwell construction result, Eq. (\ref{vm}). The critical metallic fraction $v_c$, which corresponds to the critical pressure $P_c$, was similarly calculated.
} 
\label{fig:scaling} 
\end{figure}

{\it Percolative conduction} --
The measured resistance corresponding to each transport data point, indicated by the red dots in Fig. \ref{Phase}, is shown in Fig. \ref{fig:scaling}.
The resistance  shows percolative behavior characteristic of an inhomogeneous phase consisting of interspersed metallic and insulating puddles. 
Starting from an insulator at ambient pressure, the inhomogeneous phase sets in beyond  $P \sim$  3 GPa, when the incipient metallic phase  begins to appear and increases with pressure. Conducting transport occurs beyond $P_c \sim 32$ GPa, when the volume fraction of the metallic region exceeds the percolation threshold, roughly $v_c \approx 0.29$ \cite{Note}. At a much larger pressure $P_M$
(theory predicts $P_M \sim 81$ GPa as seen from Fig. \ref{fig:EvsV}), the system would become a homogeneous single  metallic phase; however, $P_M$ is larger than our maximum pressure of 54 GPa and was not experimentally reached.     
The
Raman data\cite{Baldini} show the presence of a mixture of distorted and undistorted regions across the MIT, specifically, up to the highest measured pressure of 34 GPa, while a remarkable decrease of the intensity-noise ratio in the Raman data  at 32 GPa is a spectral signature of the onset of the MIT.

The measured resistance, presented in Fig. \ref{fig:scaling},  is described very well  by the standard percolation scaling laws for the metal-insulator composites,
viz.,
\begin{eqnarray}
R =\begin{cases}
R_1  (v-v_c)^{-t}  &    v > v_c \ ( \text{metallic regime)} \\
R_1^u R_2^{1-u}  &    v = v_c \   (\text{percolation threshold})\\
R_2  (v_c-v)^s  &     v < v_c \ ( \text{insulating regime)},
\end{cases}
\label{critical}
\end{eqnarray}
where $v$ again is the metallic volume fraction, and  $t = 1.6 -2.0$,  $s = 0.7 - 1.0$,
and $u = t / (s+t)$ are universal critical exponents for three-dimensional percolation\cite{Stauffer, Bergman, Nan}.
Our transport data (Fig. \ref{fig:scaling}) was fitted to  Eq. (\ref{critical}) by first computing the volume $V$ for a given pressure using  
the equation of state\cite{Loa} and then by finding the corresponding $v $  from Eq. (\ref{vm}).
The 
fitted critical exponents $t$ and $s$ (values listed in the Fig. \ref{fig:scaling} caption) are close to the theoretical exponents for 3D percolation,
and the sigmoid shape of the transport curve closely resembles the same for the composite media\cite{Nan}.

{\it The GEM equation for composites} -- Although a wide range of experimental results for conductor-insulator percolating systems and computer simulations can be fitted with the classic percolation equations expressed in Eq. (\ref{critical}), these equations are valid only in the limits, where the conductivity of the metallic fillers tends to infinity, while  the interspersed insulting matrix
is a perfect insulator with the conductivity tending to zero. This is satisfied quite well in our case, as justified 
{\it a posteriori} from the fitted resistance ratio $R_2 / R_1 \approx 3 \times 10^4$ (see Fig. \ref{Phase} caption). 
In many composites, this condition is not satisfied quite so well. 
For these cases, McLachlan et al.\cite{McLachlan} have proposed a phenomenological equation that has been successfully used
to fit the conductivity data of such composites. 

This so-called general effective medium (GEM) equation is in the
form of an implicit equation for the resistance $R (v)$ as a function of the metallic volume fraction,
which reads
\begin{equation}
\frac {  (1-v) (R^{1/s}  - R_2^{1/s} )   } {R^{1/s} + A R_2^{1/s}}   + 
\frac {  v (R^{1/t}  - R_1^{1/t} )   } {R^{1/t} + A R_1^{1/t} }
=0,
\end{equation}
where $A = (1-v_c)/v_c $ and $R_1$ and $R_2$ are, again, the resistances of the conductor and the insulator, respectively. This equation remains valid if the resistances are replaced by the corresponding resistivities.
It can be easily verified  that this single two-exponent percolation equation continuously interpolates between the three percolation equations in Eq. (\ref{critical})  and it reduces to a normalized form of each of them in the limits, $R_1 \rightarrow 0$ and $R_2 \rightarrow \infty$.  In the crossover regime $v \approx v_c$ (more specifically, 
$  | v - v_c| <   (R_1 / R_2 )^{ 1 / (t+s)}   $), it reduces to the middle line of Eq. (\ref{critical}).  We were able to fit our
resistance data with this equation as well, which provided a single continuous curve, with the four fitting parameters $R_1$, $R_2$, $t$, and $s$. This fitting yielded 
very similar values to the parameters reported in Fig. \ref{fig:scaling}, which were obtained by fitting the resistivity data to Eq. (\ref{critical}) in the limiting regions away from the critical region.

%
%: Temperature Dependence

\begin{figure}[t]
\includegraphics[angle=0,width=0.8 \linewidth]{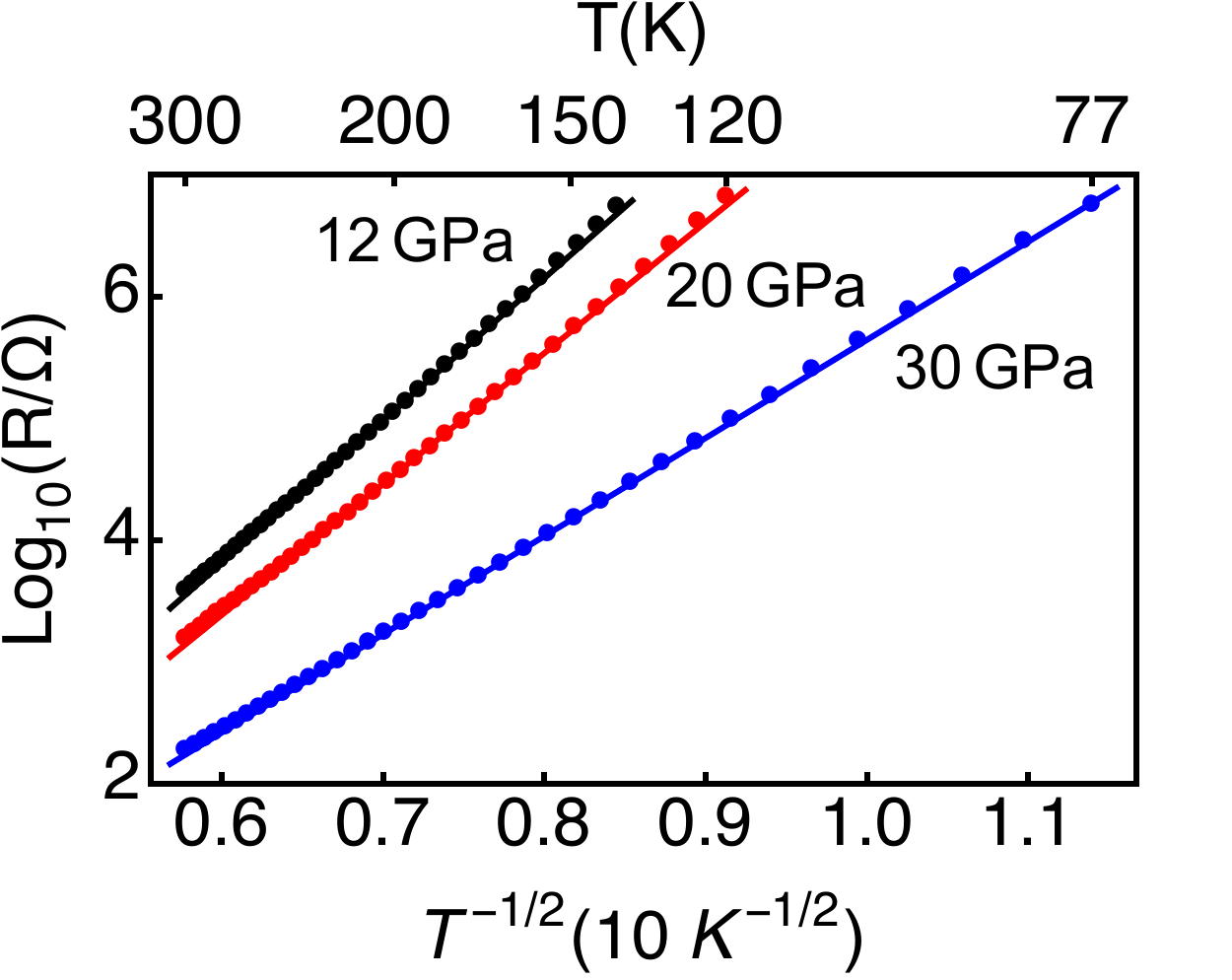}
\caption{(Color online)  
The  measured temperature dependence of resistance  on  the insulating side of the MIT 
at three different pressures. 
} 
\label{fig:Temp}
\end{figure}

{\it Temperature Dependence} --
The temperature dependence of the resististance in the insulating regime is shown in
Fig. \ref{fig:Temp}, which follows the Efros-Shklovskii variable range hopping (VRH)  behavior\cite{Efros-original, Shklovskii2012} 
\begin{equation}
R = R_0 \exp [ (T_0/T)^{1/2}],
\end{equation}
observed in a variety of granular materials\cite{Efetov}, where non-percolative metallic puddles (metallic fraction below the percolation threshold) are surrounded by  insulating material.

%:Fig.6
\section{Conclusion}

In conclusion, we studied the metal-insulator transition  in LMO under pressure using the Gutzwiller solution of a model Hamiltonian containing correlation and Jahn-Teller effects and high-pressure transport measurements. Our main result is that the MIT is driven by a combination of the correlation and Jahn-Teller effects, and it is percolative in nature, which is fundamentally  different from the standard Mott-Hubbard transition. In the present case, the MIT occurs due to percolative conduction in a mixed phase consisting of interspersed metallic and insulation regions, while in the Mott-Hubbard transition, conduction occurs due to the sudden change of the ground state of the system with some parameter, with the system maintaining a homogeneous, single phase across the MIT. The theory work showed that the system has a propensity for phase separation when volume is compressed, where the system separates into a metallic part and an insulating part separated by a single phase boundary.
 However, rather than the two parts forming two separate regions, they are interspersed among each other on the nanoscale in the experiment,  thereby forming a mixed or an inhomogeneous phase (nanoscale phase separation). The exact reasons for this is unknown, but effects such as Coulomb interaction between the two parts or kinetic reasons have been proposed in the literature as discussed in the text.

The measured high-pressure resistance followed the percolation scaling laws both as a function of temperature and pressure, establishing the percolative nature of the metal-insulator transition. As pressure is applied on LMO, an insulator at $ P = 0$, the metallic region begins to form around $P \sim 3$ GPa, with the metallic  fraction gradually growing with pressure and eventually forming a conducting network beyond the percolation threshold, which occurs at $P_c \approx 32$ GPa.
 Thus, while the MIT is sharp, caused by the onset of the percolative conduction, there is no such sharp change in the metallic volume fraction, which grows continuously across the MIT. 
In turn, since the metallic region contains undistorted JT octahedra, the average lattice distortion 
also changes continuously across the MIT as seen in  the Raman data. 
The percolative MIT may be more common place in the oxide materials than is currently thought and needs further study, both from the viewpoints of fundamental science as well as of potential  applications in oxide electronics.
% and showed that both electron-electron  and the electron-lattice interactions play important roles in the transition. 
%The MIT is very different from the standard Mott-Hubbard type, where the entire crystal changes from metallic to insulating phase. 
% in the sense that it is not a transition from a fully insulating to a fully metallic phase as happens for many solids under pressure. 
%In the present case, starting from the insulating phase at ambient pressure, the metallic domains begin to appear around 3 GPa and the system shows metallic conduction, only when the fraction of the metallic domains, growing with pressure, forms a percolating conducting network,  beyond  $P_c \approx 32$ GPa, when the metallic volume fraction exceeds the percolation threshold $v_c \approx 0.29$. 
%

\begin{acknowledgements}
We thank P. Schlottmann and B. Shklovskii   for helpful discussions. SS and MS were supported by the U.S. Department of Energy, Office of Science,
under Award    No. DE-FG02-00ER45818. MB was supported as part of the Energy Frontier Research in Extreme Environments Center (EFree), an Energy Frontier Research Center funded by the U.S. Department of Energy, Office of Science under Award No. DE-SC0001057.
\end{acknowledgements}

$^\dagger$ Present Address: Instituto de Ciencia de Materiales de Madrid, CSIC, Cantoblanco, E-28049 Madrid, Spain.

%:References

\end{document}